# Scaling indicator and planning plane: an indicator and a visual tool for exploring the relationship between urban form, energy efficiency and carbon emissions

Fouad Khan[1] (corresponding author) and Laszlo Pinter[2]

[1]Central European University and WWF Luc Hoffmann Institute, A-202, Block 7, Gulistan-e-Johar, Karachi, Pakistan; Email: fouadmkhan@gmail.com, Phone: +45 91626381, Fax: +92 (51) 9090141

[2]Professor - Department of Environmental Sciences and Policy, Central European University, Nador 9, 1051 Budapest, Hungary & Senior Fellow and Associate - International Institute for Sustainable Development, Winnipeg, Canada; Email: lpinter@iisd.ca

**Highlights**

- A scaling indicator for cities is defined and calculated for 58 US cities using census data
- The scaling indicator is correlated to greenhouse emissions and gasoline sales
- A spatial planning tool is proposed to incorporate analysis of scaling into urban development planning

**Abstract**

Ecosystems and other naturally resilient systems exhibit allometric scaling in the distribution of sizes of their elements. In this paper we define an allometry inspired scaling indicator for cities that is a first step towards quantifying the resilience borne of a complex systems' hierarchical structural composition. The scaling indicator is calculated using large census datasets and is analogous to fractal dimension in spatial analysis. Lack of numerical rigor and the resulting variation in scaling indicators -inherent in the use of box counting mechanism for fractal dimension calculation for cities- has been one of the hindrances in the adoption of fractal dimension as an urban indicator of note. The intra-urban indicator of scaling in population density distribution developed here is calculated for 58 US cities using a methodology that produces replicable results, employing large census-block wise population datasets from the 2010 US Census 2010 and the 2007 US Economic Census. We show that rising disparity – as measured by the proposed indicator of population density distribution in census blocks in metropolitan statistical areas (using US Census 2010 data) adversely affects energy consumption efficiency and carbon emissions in cities and leads to a higher urban carbon footprint. We then define a planning plane as a visual and analytic tool for incorporation of scaling indicator analysis into policy and decision-making.

# 1. Introduction

It has been shown that the hierarchical organization in ecosystems makes them more stable and less sensitive to damage from environmental disturbances (Jørgensen and Nielsen, 2013). The mechanisms underlying this 'resilience' that originates from the system 'form', have been a subject of study since the earliest analyses of systemic risk were undertaken for anthropogenic complex systems (Perrow, 1984). Higher de-coupling between system elements (or niches for ecosystems) and higher functional redundancy have been identified as factors that contribute towards making a system more adaptable and hence more resilient to shocks. Post 2008, a growing body of literature has also explored the role of these factors in making economic systems more or less resilient (Taleb, 2012). This paper explores how the city as another anthropogenic complex system can be analyzed for the presence, absence or degree of this resilience within its structure. The hierarchical organization that lends ecosystems their resilience expresses itself structurally in the form of very specific scaling. What this means is that in such systems, the design elements are distributed at various scales such that the number of elements $p$, at each scale $x$ are related according to the equation $px^m$ = constant (Salingaros and West, 1999) where $m$ is the exponent of the power law, also called the fractal dimension. Like the teeth along the edge of a toothed leaf or the orbits of moons and planets, similar design elements repeat themselves at different scales and also on the same scale. Natural complexity emerges out of a repetition of design algorithms with slight variations or anomalies or mutations for each repetition and at each varying scale. In other words, typically these systems are not naturally inclined to have aberrantly sized elements and the number of component elements decreases as the scale to which the element belongs increases in size. The bigger an element is, the lesser its population in the system (Parrott, 2010; Salingaros and West, 1999; West and Brown, 1997, 2004; West et al., 1999).

In architecture and urban planning there has been an emerging body of work rediscovering the significance of form and scaling in urban planning especially within the new urbanism movement (Batty and Longley, 1994; Benguigui and Czamanski, 2004; Bettencourt et al., 2010; Coward and Salingaros, 2004; Salingaros and West, 1999; Shen, 2002). These form analyses have taken into account the complex nature of urban systems and identified fractal dimension as an indicator of note, both as a measure of scaling and space filling within the city. It has also been shown that on a greater scale, similar scaling characteristics can be attributed to the distribution of human population in general, with cities having predictable socioeconomic and infrastructural parameter values based on their size (Bettencourt et al., 2010; Hern, 2008).

Despite progress at the conceptual level, the role of scaling characteristics in understanding the role of urban form hasn't been universally recognized. Continued skepticism towards the significance of form in urban planning (Echenique et al., 2012) however, often fails to take into account the complexity of urban systems while analyzing form. One of the reasons for that is because scaling indicators such as fractal dimension, are often not easy to calculate with replicability and reliability. A case in point is the box counting method that has been traditionally used for estimating scaling indicators for cities (Batty and Longley, 1994; Benguigui et al., 2000; Hern, 2008; Shen, 2002). This usually involves overlaying a grid on a digitized map of the city and then counting or estimating the covered or relevant populated area within each box of the grid. The count is then binned into classes according to increasing size or increasing number of boxes (having count within the class range) within each class. The scaling indicator is estimated by plotting a log-log graph of the count range against the number of boxes falling within that count range; the slope of the resulting trend-line is the exponent of the power law or our scaling indicator of concern for the distribution of sizes of elements. The method is prone to varying results given the size of the box and the resolution of the map or image. Of course this lack of replicability means that the indicator does not meet a fundamental criterion for good indicator development and represents a constraint on its usefulness for policy (Kandziora et al., 2013; Pintér et al., 2012).

In this paper, we present a more rigorous method for estimating a scaling indicator for cities that can produce replicable results. The new method would allow a more reliable representation and analysis of urban form, and an indicator of space filling within the city can be developed that is cognizant of the complex nature of the city. The paper goes on to show how urban form, once analyzed in this manner, does indeed influence sustainability attributes such as gasoline consumption within the city. We also suggest that further research into more reliable scaling-based indicators such as the one proposed is warranted and could result in discovering new relationships between spatial structure and environmental performance with significant relevance for policy. Finally we propose a visual representation of the new indicator called planning plane to incorporate analysis of scaling into policy for urban sustainable development.

## 2. Materials and Methods

To implement our data intensive method for the estimation of a fractal dimension based scaling indicator, data on US population by census blocks is downloaded from the US Census Bureau website (US Census Bureau, 2010). A census block is a small unit roughly congruent to a neighbourhood block. As such, the assumption that the housing type within the census block is largely homogenous should hold. The data is downloaded for Metropolitan Statistical Areas (MSAs) which are census designated places that take into account the network of economic, industrial and commercial activity. So if a suburb has most of its financial linkages to a metropolitan area, the corresponding MSA would include the suburb as part of the MSA. The MSA is selected as the smallest unit of analysis for the study.

The census blocks for each city are first sorted according to increasing population density and then binned in ten classes using k-means clustering (Lloyd's algorithm) along the population density spectrum (Khan, 2012). The population density and area covered is then calculated for the ten classes. The fractal dimension based scaling indicator is calculated by plotting the inverse of population density against the area covered by housing of that density. Once plotted on log-log scales the resulting slope of the trend-line would be the exponent of power law or the proposed fractal dimension based scaling indicator of the distribution of population densities within the city.

To derive our exponent we first start with the formula for the box-counting dimension, as expressed by **Equation 1** (Salingaros and West, 1999). According to the box-counting method a grid of 'boxes' is layered on a map of the city, that divides the spatial spread of the city into different populated areas, each with a different land use coverage.

$$D = \frac{\log N_x}{\log(\frac{1}{x})} \quad (1)$$

Where,

$D$ = box counting dimension

$x$ = certain percentage (or range of percentages) of area of the box covered by land use

$N_x$ = Number of boxes falling within range $x$

Instead of a map we have an extensive data set of the distribution of urban population by census-blocks. Accordingly, instead of overlaying a grid of 'boxes' on a map, we will split the population into virtual boxes, each covering an area of one kilometre square. So in our methodology the 'box' of the box-counting method is any given one-kilometre square region of the city.

The next step is to establish the frequency distribution of population intensity across the boxes. In box counting method this is done by counting all the boxes that fall within a certain range of land use coverage; say 2 out of 20 boxes have between forty to fifty percent of their area covered by urban land use. This is designated by the term $N_x$ in Equation 1. For our methodology the congruent count will be the number of km² boxes that fall within a certain population density range; say 8 km² of the city has population density between 100 and 200 people per km². The comparison of these two methodologies is shown in **Figure 1**. In the proposed method, an area of one km² is analogous to what is defined as the 'box' in box counting method and clustering is on the basis of population density instead of percentage of area of box covered by land use.

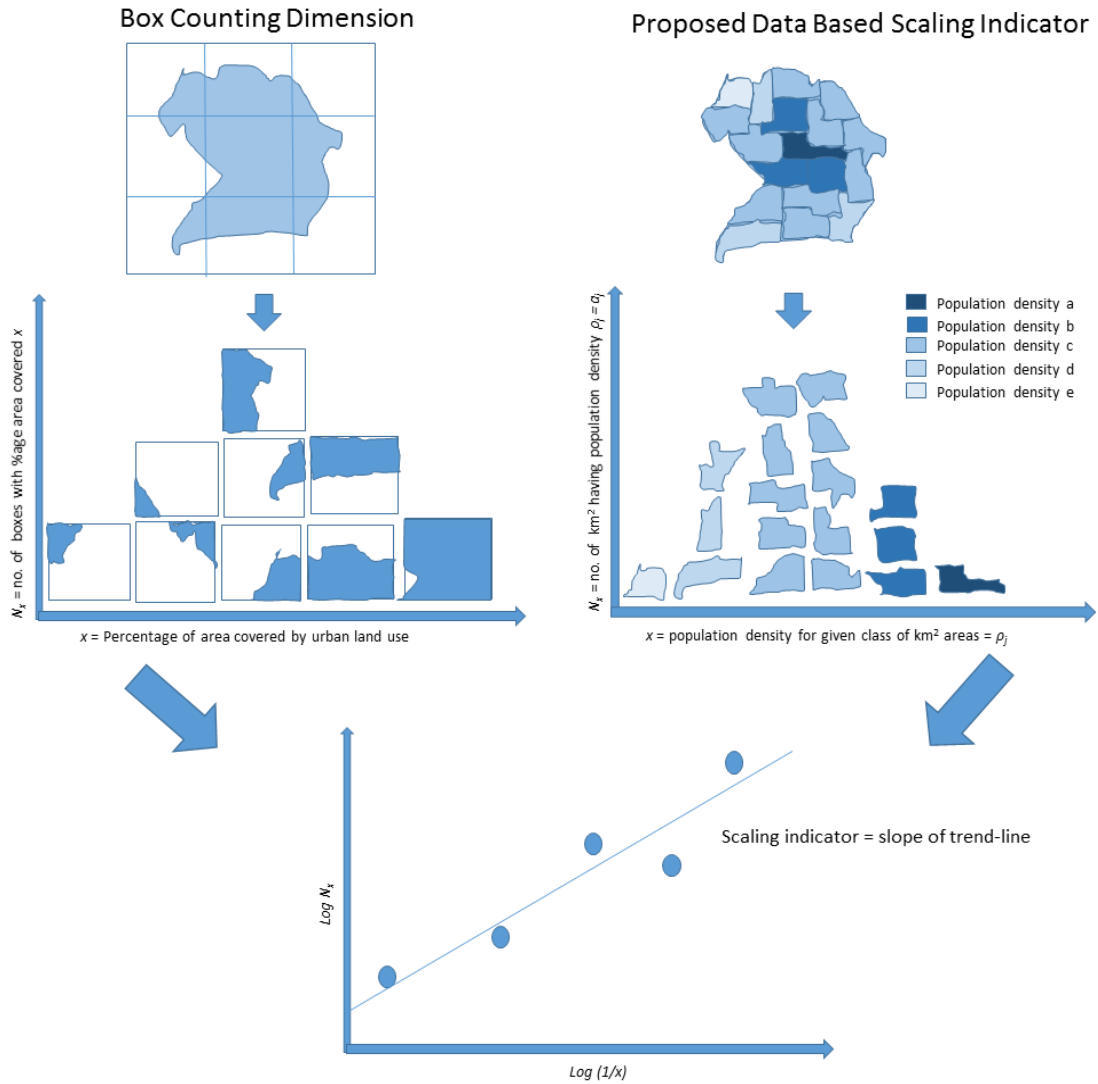

**Figure 1. Comparison of image/map based box-counting method to proposed numerical data-based methodology**

So if, $a_i$ = *area of block i*, where *i = 1 to n*, and *n = no. of blocks*, and $p_i$ = *population in block i*, then population density in block *i* can be given as;

$$\rho_i = \frac{p_i}{a_i}$$

Now if the list of blocks is sorted according to increasing population density $\rho_i$ such that increasing index *i* indicates blocks of increasing density and blocks are clustered in *c* number of classes such that the blocks falling within each class *j* have the population density range $\rho_{il} \leq \rho_i < \rho_{iu}$, where $\rho_{il}$ = *lower population density bound of class* and $\rho_{iu}$ = *upper population density bound of class*, then the total area for any class *j* can be given as;

$$a_j = \sum_{i=l}^{u} a_i$$

where, $l = i$ when $\rho_i = \rho_{il}$ and $u = i\text{-}1$ for $\rho_i = \rho_{iu}$ for any given class *j*.

Similarly the population and population density for the class *j* can be written respectively as;

$$p_j = \sum_{i=l}^{u} p_i$$

$$\rho_j = \frac{p_j}{a_j} = \frac{\sum_{i=l}^{u} p_i}{\sum_{i=l}^{u} a_i}$$

Since for our scaling exponent, the no. of boxes $N_x$ falling within a certain range *x*, is simply the number of square kilometres $a_j$ falling within the population density range for a given class *j*, i.e. $\rho_{il} \leq \rho_i < \rho_{iu}$, $N_x$ can be written simply as;

$$N_x = a_j = \sum_{i=l}^{u} a_i$$

Since we are clustering the blocks into classes based on population density, the parameter used to measure scaling, i.e. *x* in Equation 1 would be the population density for a given class $\rho_j$, for which we count the kilometre squares or measure the area $N_x = a_j$. Therefore;

$$x = \rho_j = \frac{p_j}{a_j} = \frac{\sum_{i=l}^{u} p_i}{\sum_{i=l}^{u} a_i}$$

The scaling indicator for our calculation, say $D_s$ can now be expressed as;

$$D_s = \frac{\log N_x}{\log(\frac{1}{x})} = \frac{\log a_j}{\log(\frac{1}{\rho_j})}$$

Using Richardson-Mandelbrot slope (Mandelbrot, 1967) now the value of $D_s$ will be calculated by plotting $a_j$ and $\rho_j$ on log-log scale and estimating the slope of the regression line as shown in **Figure 2**. As $a_j$ is measured in km$^2$ and $\rho_j$ is measured in km$^{-2}$, $D_s$ is a dimensionless quantity.

A total of 58 cities were included in final analysis. The following heuristic was followed to arrive at the final list of cities.

1. A set of ten cities were initially selected to run a test study. These cities were selected by dividing the entire set of MSAs into five groups based on city size, assigning random numbers to cities and selecting the first ten insuring that each of the five city size groups as well as maximum variation in states, percentage change in population over the last ten years, urban topography and climate could be captured.

The remaining cities were listed alphabetically to arrive at a certain pseudo-randomness. The first sixty-eight cities were selected for analysis. Four cities lying in multiple states or sharing counties with different cities were ignored. Hence the scaling indicator was calculated for seventy-four US cities in total (including the first ten). Out of these thirteen were excluded from the analysis because the data showed a difference between the geographically calculated population numbers and those reported in US Census data for the metro area on a cumulative basis because of privacy protection concerns in reported census data. For three cities, the data for energy usage was not available due to the same reason. This left a dataset of fifty-eight cities for final analysis.

To establish the veracity and utility of the scaling indicator of disparity, its correlation with energy consumption efficiency was studied. Gasoline station sales data was taken as a proxy for gasoline usage. Data on sales at gasoline stations within the MSAs was downloaded from the US Economic Census 2007 website (US Census Bureau, 2007). The gasoline station sales data is from 2007 and sales data for 2010 is not available. However, for the year 2010 data is available for gasoline station payrolls. The sales data is extrapolated for 2010 using the percentage change in total payroll from 2007 to 2010. Additionally, the correlation between disparity measured using the scaling indicator and carbon emissions per capita is studied. The data on $CO_2$ emissions is for the year 2008 and downloaded from Arizona State University's Vulcan Project (The Vulcan Project, 2012). The emissions considered for the analysis were only for road transportation. The correlation between scaling indicator and energy consumption was studied using least-squares method.

3. **Theory**

As can be seen in **Figure 1**, the box counting method is based only on the spatial extent of urban areas, irrespective of any of their more detailed attributes, while the proposed indicator provides a more fine-tuned quantitative measure by taking population density into account. In that sense the proposed scaling indicator measures the scaling of population distribution along a third dimension and sees the city as a three-dimensional fractal (with space filling in the vertical dimension accounting for higher fractal dimensions), rather than a 2-dimensional plane. If a housing unit is the basic self-similar geometry filling the urban space then multi-storey buildings with multiple housing units may be analogous to the city being a three-dimensional fractal. The scaling indicator proposed thus is a more accurate measure of how human population constitutes the city as a three dimensional space.

The indicator is also a measure of disparity within the system. Compare the cities in **Figure 2** for instance; Houston, Texas has a higher area covered by its lowest density housing and lesser area covered by its highest density housing compared to Pine Bluff, Arizona. Thus there is greater disparity between extremes in Houston, Texas compared to Pine Bluff, Arizona and thus it has much higher scaling indicator value (slope of the trend-line).

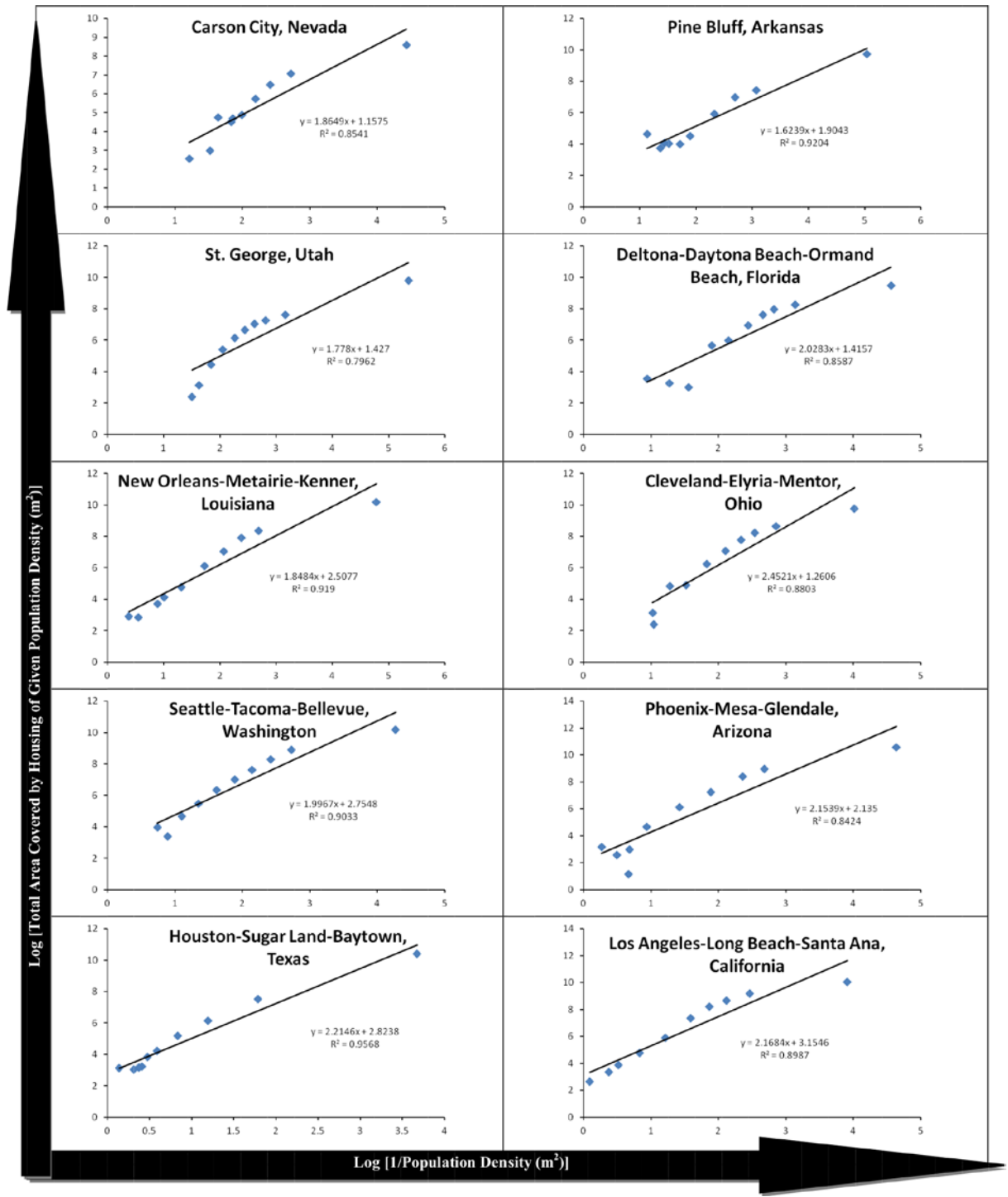

**Figure 2. Calculation of scaling indicator for ten sample cities showing how higher disparity of population density distribution leads to higher scaling indicator value**

A visualization of how this scaling indicator captures urban disparity is shown in the 'fractal spectra' developed for ten sample cities in **Figure 3.** A fractal spectrum is defined as a plotting of squares representing different areas covered by various population density blocks on an x-axis representing the specific population density. The color represents the total amount of area covered and for any given city, the change across the color spectrum represents change in the area covered by specific population density housing. Rapid change in color across a thin population density band is indicative of greater disparity.

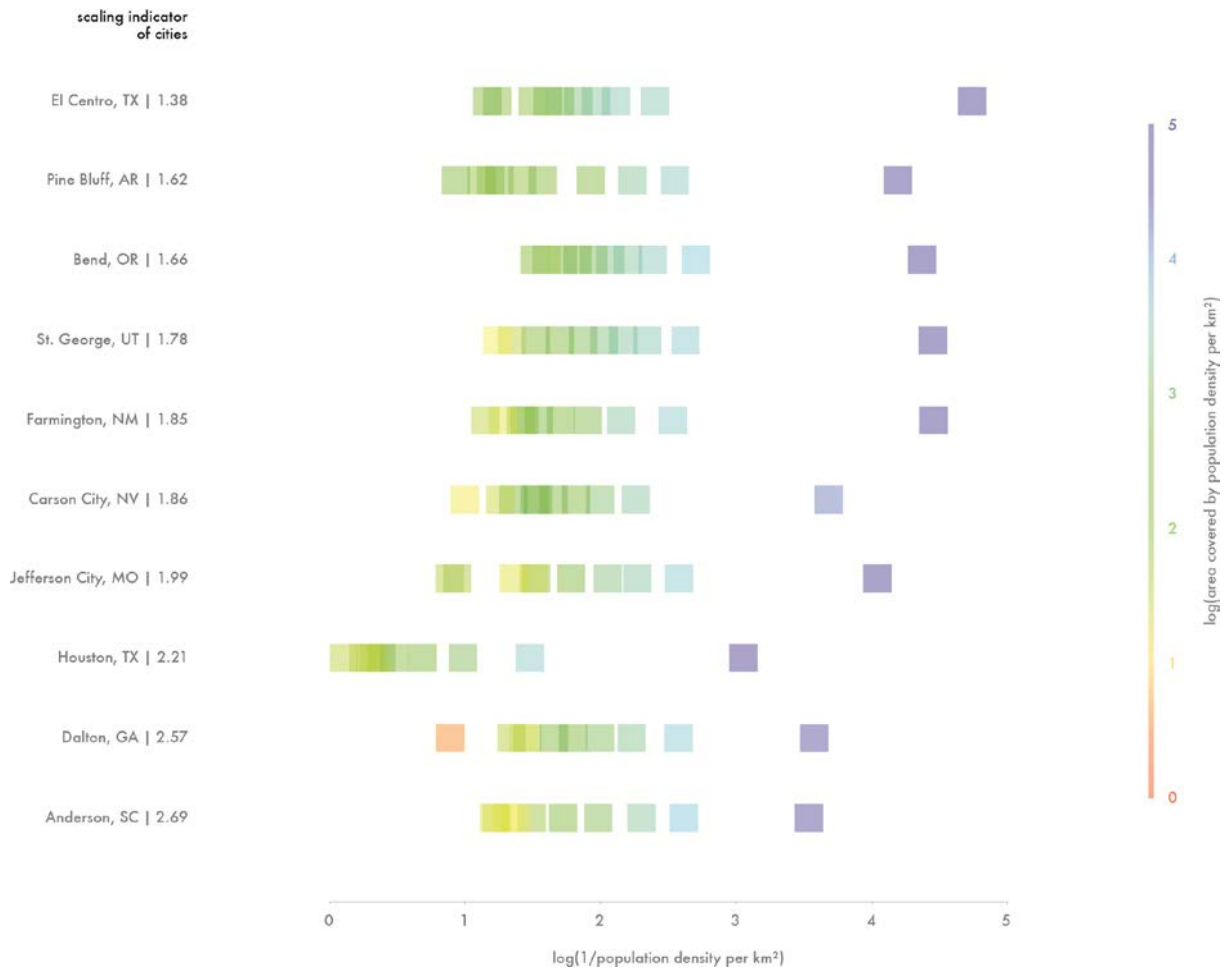

**Figure 3. 'Fractal Spectra' for cities; a plotting of colored squares (with color representing different areas covered by various population density blocks) on an x-axis representing the specific population density. Rapid change in color across a thin population density band indicates greater disparity**

## 4. Results

To establish the veracity and utility of the scaling indicator of disparity, its correlation with energy consumption efficiency was studied. The primary reason for selecting energy consumption efficiency as a dependent variable in the analysis was the significance of energy consumption towards making human development more sustainable. Both climate change and energy supply are some of the defining challenges of our time at all levels of governance and demand a lean energy diet for the future (Birol, 2013; Stocker et al., 2013). If disparity in urban or economic systems can be shown to be co-occurring with energy consumption efficiency then it becomes imperative to consider this parameter in development policy and planning.

For cities, as the scaling indicator calculated was an indicator of urban form in a manner similar to fractal dimension, the effect of the indicator disparity on gasoline usage per unit of urban area was considered. Additionally, the correlation between the measured disparity using a scaling indicator and carbon emissions per capita is studied.

As shown in **Figure 4**, as disparity in population density distribution increases in cities, gasoline usage in the city for road transportation also increases. The result of this increase can also be seen in a similar change in road transport related carbon emissions.

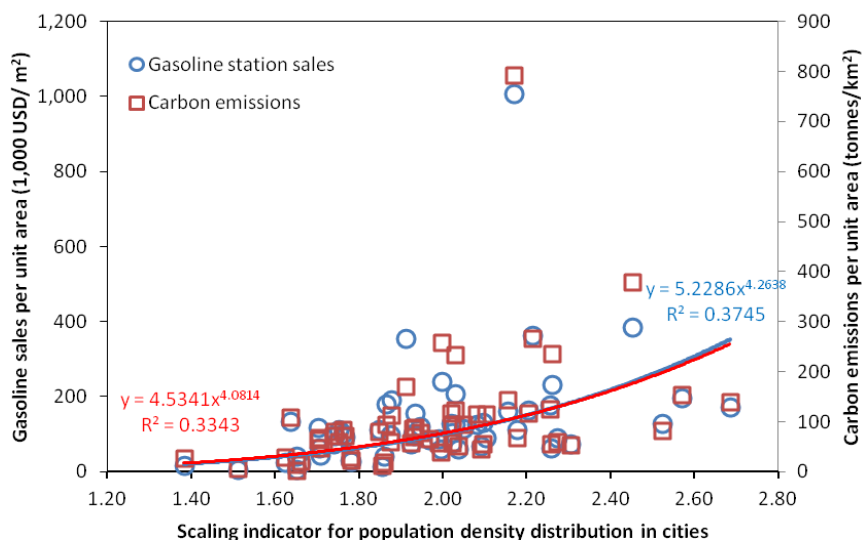

**Figure 4. Correlation between the scaling indicator, energy use and carbon emissions; both increase with increasing scaling indicator value (and thus with growing disparity in population density distribution)**

As shown on **Figure 4**, the response of energy consumption indicators to scaling based disparity indicators in non-linear. Further it can be seen that cities associated with high levels of sprawl such as Houston or Los Angeles are shown to have grater scaling indicator value and higher GHG emissions and gasoline usage.

To incorporate heuristic consideration of this variation in energy efficiency with changing scaling indicator in planning process for cities and national economies, a visual tool called planning plane is proposed here as an indicative tool for incorporating multivariate concerns in urban planning. With the increasing complexity of decision contexts, urban planning is often taking place in multi-objective settings, where the consideration of just the average value of variables is no longer sufficient. This is particularly true in cases when cities pursue high-level policy objectives such as well-being or sustainable development goals (SDG) that require the consideration of more than one key variable or the average of the variable value.. New visualization tools can ease communication of complex ideas in a way that would facilitate consideration of multiple variables. The planning plane is just such a tool in that it shows how a dependent variable changes its value based on two independent variables. The x and y axes are independent variables and color or contour can represent the dependent variable. The plane as used in this study is built using empirical datasets and spatial interpolation (ordinary kriging was used in this research). Statistical diagnostics should of course be run on the interpolation to ensure that the data available is sufficient for the construction of the planning plane.

The planning planes showing how gasoline sales per unit area and carbon emissions change with change in population density and scaling indicator of population density distribution in US cities (based on data from 58 MSAs) are shown on **Figure 5a** and **Figure 5b.** Notice that the roughness in the planes provide some opportunity to capture the non-linearity of response of the dependent variable against changes in independent variable values. The planning plane could be used in ex-post and ex-ante planning. In the ex-post context the decision-maker would obtain historic population density, GHG emission and gasoline consumption data, and calculate a sequence of planning planes for a series of time slices to assess how changes in the urban form as expressed through the spatial distribution of population affected gasoline consumption and emissions.

      To incorporate the scaling indicator in ex-ante analysis the decision maker would need to create future urban development scenarios and input new developments associated with each scenario in his map of the city and calculate the new scaling indicator value and change in population density. Using these two variables she will then plot how the energy use in the city will be affected by the new housing development or policy change, by plotting the new scaling indicator and population density values on the planning plane and comparing with previous results. If the calculation in a given scenario shows a potentially higher than expected increase in energy use, she can factor this result in the decision for letting the housing development or new policy changes proceed as proposed or develop alternatives.

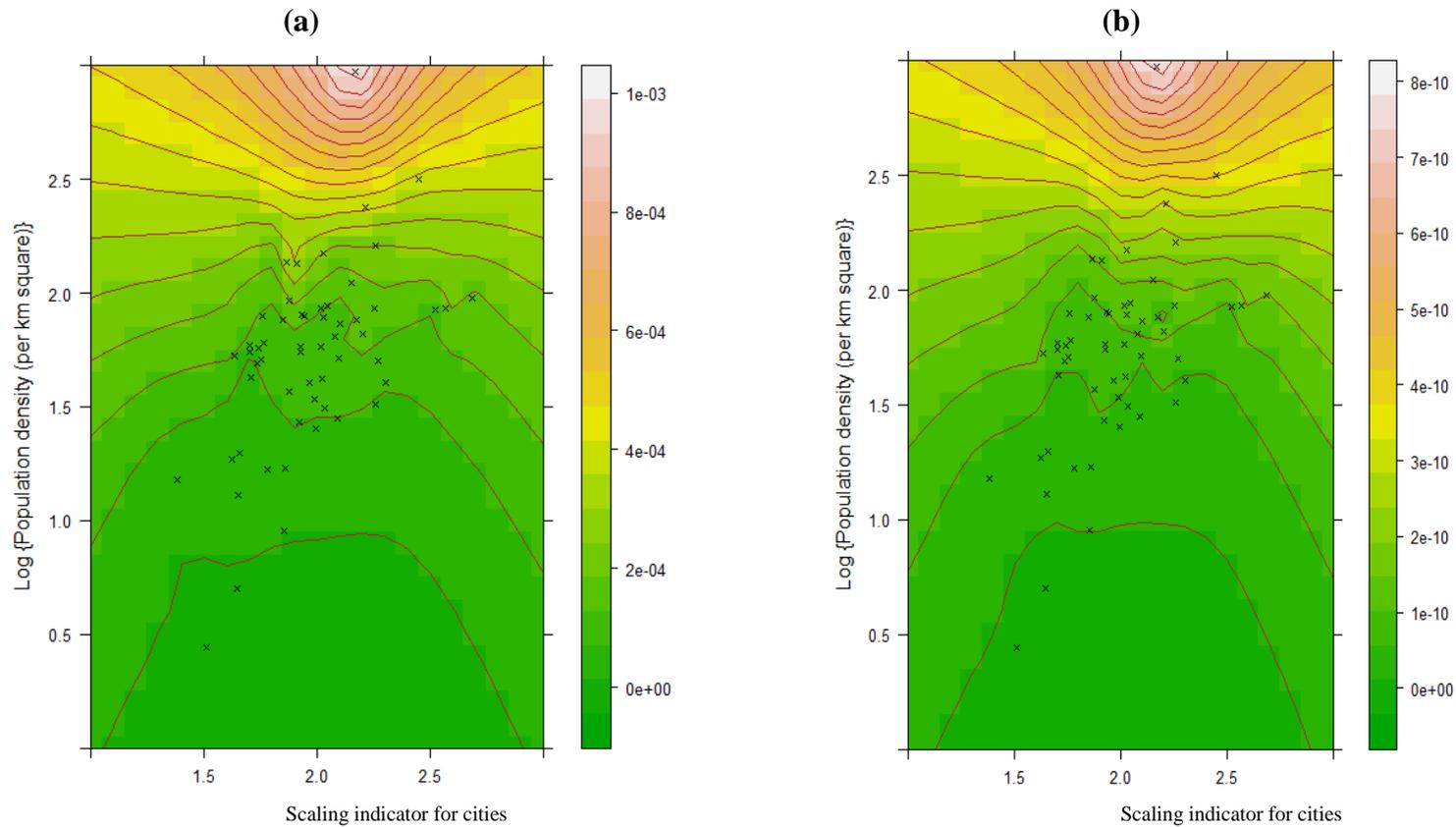

**Figure 5. Planning plane showing how (a) gasoline sales per unit area and (b) carbon emissions change with change in population density and scaling indicator of population density distribution in US cities (based on data from 58 MSAs)**

## 5. Discussions

To understand why the correlation between energy consumption and a scaling indicator exists, we need to understand what scaling implies. A higher value of the scaling indicator means that the change in distribution of certain properties across different scales is steeper. For instance for cities, a higher fractal dimension would mean that greater urban area is occupied by lower density housing compared to a city with lower fractal dimension. Higher scaling indicator values would also mean that for cities, the spread of population density (difference between minimum and maximum) is lesser or the difference between areas covered by minimum and maximum density housing is higher or both, and for economies it would mean that disparity in distribution of incomes (difference between minimum and maximum) is higher, assuming that the distribution of income follows the distribution of population density. In general, higher scaling indicator values are indicative of higher disparity in distribution of elements across the system, i.e. anomalous land use pattern with a high percentage of the land occupied by very few people or an excessively large part of the population living in overcrowded conditions.

Typically in suburban cities with extensive sprawl higher scaling indicator value implies that because of the absence of mixed land-use, i.e. concentration of functions such as commercial activity and living in specific areas, there is great disparity between population density across different blocks or areas. This would also explain why as the disparity increases as measured by the scaling indicator, energy consumption also goes up. Because of distribution of functions, people living in lower-density suburbs have to travel greater distances for functions such as work and leisure, thereby increasing energy usage.

We now try to understand this phenomenon in general system terms. Bettencourt (Bettencourt, 2013) proposed that the minimum productivity for a viable addition of a citizen to the population of a city should be equivalent to $T=\epsilon A^{(H/D)}$, where T is the energy cost associated with exploring the city fully, $\epsilon$ is cost per unit length, A is the city's land area, H is the Hausdorff fractal dimension and D is the spatial dimension (two) of the city. Since the scaling indicator for cities calculated here is analogous to the Hausdorff dimension, our work provides empirical evidence for the proposition that the energy cost associated with exploring the city fully does indeed increase with an increase in the Hausdorff fractal dimension or scaling indicator. We propose that this increase in energy consumption with increasing disparity may be owed to the general increasing energy cost of 'good regulation' in systems (Conant and Ross Ashby, 1970) as the system becomes more disparate and unequal.

A number of indicators have been developed that incorporate scaling analysis for cities, however all of them utilize spatial image or satellite data. This is the first indicator developed to use very high resolution (block wise) census data for the calculation of a scaling indicator for cities. Direct policy implications of this research can include, among others, complementing existing sustainability indicator and standards systems such as LEED-Neighborhood that do not currently have a scaling indicator incorporated into their indicator set and management system (USGBC, 2007). However, as discussed in detail earlier, this research can be viewed as part of a greater ongoing, multi-disciplinary, multi-sector effort to develop a science of sustainability with focus on low certainty and high stakes problems, such as climate policy, urbanization or societal metabolism. The fractal dimension as calculated here using a highly replicable and standardized methodology has the potential to contribute to the development of a more advanced paradigm for infrastructural and urban development planning and investment that favors resilience and sustainability already at the structural level as opposed to economic maximization only. Such a systemic indicator is needed because we are dealing with complex systems and direct input, output or process-related indicators are by themselves not adequate as a good predictor of system behavior.

To visualize an application, imagine that the city of St. George, Utah receives an application for a new housing development. To incorporate fractal dimension in the analysis the planner would input the new housing development plan in the map of the city and calculate the new fractal dimension and related population density for the proposed development as a scenario. Using these two variables the planner will then plot how energy use will be affected by the new housing development overall, by plotting the new fractal dimension and population density values on the planning plane and comparing with previous results. If there is a potential increase in energy use expected, the planner can factor this result in the decision for letting the housing development proceed as proposed or develop alternatives.

6. Conclusions

In this paper we present a scaling indicator based on extensive US Census 2010 data to calculate the scaling of population densities in 58 US cities. We notice that population densities are power-law distributed and the exponents of the power-law or scaling indicator or fractal dimension has an effect on gasoline consumption efficiency and carbon emissions in cities. Scaling indicators for urban topographies have not been calculated using such high resolution population data before. The method describes brings additional replicability to the calculation of scaling indicators for urban form analysis. We then also propose a new visual tool to facilitate inclusion of analysis of scaling indicators into policy development. We believe there is sufficient evidence here to suggest that urban disparity, measured using the fractal dimension based, data intensive scaling indicator proposed here does indeed affect energy consumption efficiency in urban systems and consideration of scaling indicators of disparity in policy analysis as well as further research in the area, is merited. A first step could be to incorporate planning planes of both mean and scaling indicators for all quantities in policy analysis.

**Acknowledgements** Paul Heinicker of University of Applied Sciences at Amsterdam is thanked for his work on developing 'Fractal Spectra'.

**Author Contributions**: Fouad Khan performed most of the analysis and contributed to the conceptualization and writing. Laszlo Pinter contributed to development of the idea and writing.

There are no competing financial interests that need declaration.

Correspondence and requests for materials should be addressed to fouadmkhan@gmail.com